\documentclass[aps,prd,showpacs,twocolumn,groupedaddress,sort&compress]{revtex4-1}

\usepackage{graphicx}
\usepackage{amsmath}

\begin{document}

\title{Bottomonium in a Bethe-Salpeter-equation study}


\author{M. Blank}
\email[]{martina.blank@uni-graz.at}
\affiliation{Institut f\"ur Physik, Karl-Franzens-Universit\"at Graz, A-8010 Graz, Austria}

\author{A. Krassnigg}
\email[]{andreas.krassnigg@uni-graz.at}
\affiliation{Institut f\"ur Physik, Karl-Franzens-Universit\"at Graz, A-8010 Graz, Austria}

\date{\today}

\begin{abstract}
Using a well-established effective interaction in a rainbow-ladder
truncation model of QCD, we fix the remaining model parameter to the bottomonium
ground-state spectrum in a covariant Bethe-Salpeter equation approach
and find surprisingly good agreement with the available experimental data
including the $2^{--}$ $\Upsilon (1D)$ state. Furthermore, we
investigate the consequences of such a fit for charmonium and light-quark
ground states.
\end{abstract}


\maketitle

\section{Introduction\label{sec:intro}}


In QCD, mesons appear as bound states of (anti)quarks and gluons. The
bottomonium system, in particular below the $B\bar{B}$ threshold, is a
prototype for the successful description of a meson using a straight-forward
$\bar{q}q$-picture (for reviews on the subject, see e.g. the comprehensive
compilations of the quarkonium working group \cite{Brambilla:2004wf,Brambilla:2010cs}).
Such a simple setup is expected to be generally more realistic and accurate for
heavier than for light quarks. In a much similar way, in a covariant
Bethe-Salpeter-equation (BSE) approach a simple truncation like the well-established
rainbow-ladder (RL) truncation is expected to be more accurate for heavier quarks
and their bound states. We perform a basic initial test for this hypothesis by
employing a model historically set up to describe light mesons. In this way, we
study ground-state mesons for spins $J=0,1,2$ starting with bottomonium
down to light quarks and check the possibility to arrive at a reasonable
agreement with experiment without fine-tuning. The latter point is important, since
a general result or trend must be visible before one optimizes or fine-tunes
the available model parameters.

The paper is organized as follows: in Secs.~\ref{sec:model} and \ref{sec:equations} we review the
necessary details of the approach and the interaction. Section \ref{sec:bottomonium}
deals with the bottomonium ground states, followed by an investigation
of the consequences of our parameter choice for lower quark masses in Sec.~\ref{sec:lower}.
We conclude and present an outlook in Sec.~\ref{sec:co}.
All calculations have been performed using Landau-gauge QCD in Euclidean momentum space.

\section{Mesons in an RL model of QCD\label{sec:model}}

We employ QCD's Dyson-Schwinger-equations (DSEs) (see, e.g.~\cite{Fischer:2006ub,Roberts:2007jh}
for recent reviews) coupled with the quark-antiquark Bethe-Salpeter equation (BSE). The latter is
the covariant bound-state equation for the study of mesons in this context \cite{Hu:1966ab,Hu:1966bs,Smith:1969az},
and analogously one can use a covariant approach to baryons in both a quark-diquark picture
(e.g.~\cite{Eichmann:2007nn,Nicmorus:2008vb,Eichmann:2010je} and references therein)
or a three-quark setup \cite{Eichmann:2009qa,Eichmann:2009zx}.

Numerical hadron studies such as the present one make a truncation of this infinite tower of coupled and in
general nonlinear integral equations necessary. Herein, we use the so-called rainbow-ladder (RL) truncation,
which is well-established as a tool for modeling hadron physics in QCD. In particular, it is better-suited
as an approximation to the full set of equations the higher the quark mass becomes,
see e.g., \cite{Bhagwat:2004hn,Eichmann:2008ae,Eichmann:2008ef}. Related concrete results on the heavy-quark
domain and -limit of Coulomb-gauge QCD have become available recently \cite{Popovici:2010mb,Popovici:2011yz,Popovici:2011wx}.

The RL truncation is simple yet offers the possibility for sophisticated model studies of QCD within the DSE-BSE context,
since it satisfies the relevant (axial-vector and vector) Ward-Takahashi identities
(see e.g.~\cite{Maskawa:1974vs,Aoki:1990eq,Kugo:1992pr,Bando:1993qy,Munczek:1994zz,Maris:1997hd,Maris:1999bh,Maris:2000sk}).
Regarding the meson spectrum a generally accurate description on the basis of a purely phenomenologically oriented model
is conceivable. However, to increase the predictive power of the model it is advisable to reduce the
free parameters in such a model, or more precisely in the effective model interaction, as much as possible.

The expectation in such a situation is that the description of light mesons will not be as accurate, which
is a consequence of the fact that additional terms in the dressed quark-gluon vertex, which are omitted
in RL truncation, cannot be successfully mimicked by a simple parametrization of the effective interaction
such as the one used here. In the present study however, this is a defect we are willing to accept in
order to check the validity of our assumptions about the heavy-quark domain.
Our restrictions are further justified by meson studies beyond RL truncation
(see, e.g., \cite{Williams:2009ce,Chang:2010jq} for relevant references) that
have confirmed effects from correction terms, but at the same time shown both the numerical
complexity of such investigations as well as the uncertainty of the size of even further corrections.

In RL truncation the axial-vector Ward-Takahashi identity dictates that the rainbow-truncated integral-equation kernel of the quark
DSE corresponds to the ladder-truncated integral-equation kernel of the quark-antiquark BSE as given below.
The identity is crucial to correctly realize chiral symmetry and its dynamical breaking in the model
calculation from the very beginning. As the most prominent result, one satisfies Goldstone's theorem \cite{Munczek:1994zz}
and obtains a generalized Gell-Mann--Oakes--Renner relation valid
for all pseudoscalar mesons and all current-quark masses \cite{Maris:1997tm,Holl:2004fr}. This relation
can also be checked numerically and is satisfied at the per-mill level in our calculations.

We start out from a model setup defined in Ref.~\cite{Maris:1999nt} and given in detail below, which has since been
successfully applied to many in particular pseudoscalar- and vector-meson properties in recent
years (see e.g.~\cite{Roberts:2007jh,Krassnigg:2009zh} for comprehensive bibliographies). Of
interest are electromagnetic hadron properties
\cite{Maris:1999bh,Maris:2005tt,Holl:2005vu,Bhagwat:2006pu,Eichmann:2011vu},
strong hadron decay widths \cite{Mader:2011zf}, valence-quark distributions of pseudoscalar mesons
\cite{Nguyen:2010ph,Tandy:2010dw,Nguyen:2011jy},
a study of tensor mesons \cite{Krassnigg:2010mh} and an
exploratory application of this model to the chiral phase transition of QCD at finite
temperature \cite{Blank:2010bz}. Of immediate interest are the recent steps to the
successful numerical treatment of heavy quarks in this particular model setup
\cite{Maris:2006ea,Krassnigg:2009zh,Souchlas:2009ph,Souchlas:2010zz}.

\section{Quark DSE and meson BSE\label{sec:equations}}

\begin{figure}
\includegraphics[width=\columnwidth,clip=true]{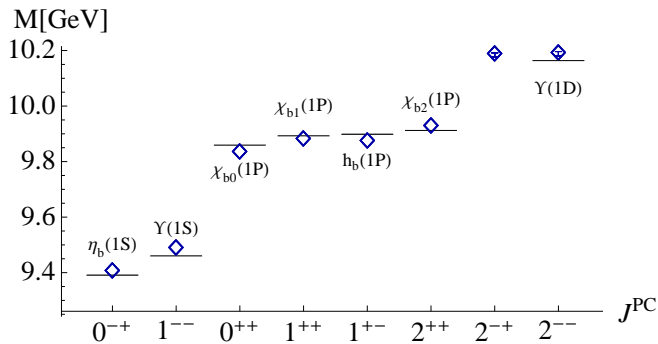}%
\caption{(Color online) The bottomonium ground-state spectrum compared to experimental
data.\label{fig:bottomonium}}
\end{figure}

In RL truncation one considers a meson with total $q\bar{q}$ momentum $P$ and relative $q\bar{q}$
momentum $q$ by consistently solving the homogeneous, ladder-truncated $q\bar{q}$ BSE
\begin{eqnarray}\nonumber
\Gamma(p;P)&=&-\frac{4}{3}\int^\Lambda_q\!\!\!\!\mathcal{G}((p-q)^2)\; D_{\mu\nu}^f(p-q) \;
\gamma_\mu \; \chi(q;P)\;\gamma_\nu \, \,,\\\label{eq:bse}
\chi(q;P)&=&S(q_+) \Gamma(q;P) S(q_-)\,,
\end{eqnarray}
on the one hand, where the semicolon separates four-vector arguments, and the rainbow-truncated quark DSE
\begin{eqnarray}\nonumber
S(p)^{-1}  &=&  (i\gamma\cdot p + m_q)+  \Sigma(p)\,,\\\label{eq:dse}
\Sigma(p)&=& \frac{4}{3}\int^\Lambda_q\!\!\!\! \mathcal{G}((p-q)^2) \; D_{\mu\nu}^f(p-q)
\;\gamma_\mu \;S(q)\; \gamma_\nu \,
\end{eqnarray}
on the other hand.

The solution of the BSE is the Bethe-Salpeter amplitude (BSA) $\Gamma(q;P)$ which,
combined with two dressed quark propagators $S(q_+)$ and $S(q_-)$ gives the
``Bethe-Salpeter wave function'' $\chi(q;P)$. Note that in the case of spin $J>0$ the
BSA carries $J$ open Lorentz indices (for details see \cite{Krassnigg:2010mh}),
which are omitted here for simplicity together with
the BSA's Dirac and flavor indices. The factor
$\frac{4}{3}$ comes from the color trace, $D_{\mu\nu}^f(p-q)$ is the free gluon
propagator, $\gamma_\nu$ is the bare quark-gluon vertex, $\mathcal{G}((p-q)^2)$
is the effective interaction specified in detail below, and
the (anti)quark momenta are $q_+ = q+\eta P$ and $q_- = q- (1-\eta) P$.
$\eta \in [0,1]$ is referred to as the momentum partitioning parameter and is usually
set to $1/2$ for systems of equal-mass constitutents, which we do as well.
$\int^\Lambda_q=\int^\Lambda d^4q/(2\pi)^4$ represents a translationally invariant
regularization of the integral, with the regularization scale $\Lambda$ \cite{Maris:1997tm}.
$\Sigma(p)$ denotes the quark self energy and $m_q$ the current-quark mass.
The solution for the quark propagator $S(p)$ requires a renormalization procedure, the details of which
can be found together with the general structure of both the BSE
and quark DSE in \cite{Maris:1997tm,Maris:1999nt}.

For the homogeneous, ladder-truncated $q\bar{q}$ BSE numerical solution methods have been
improved in recent years \cite{Blank:2010bp} and also cross-checked with the closely related
\begin{figure}
\includegraphics[width=\columnwidth,clip=true]{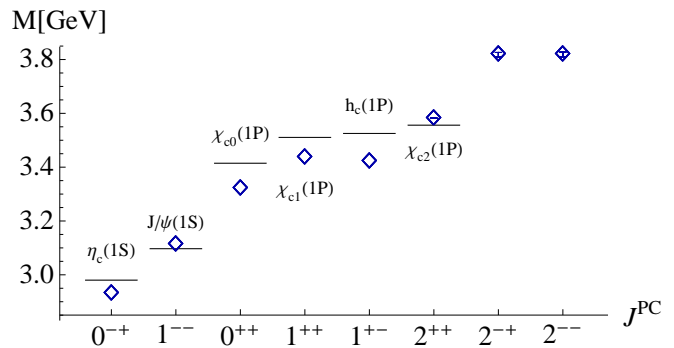}%
\caption{(Color online) The charmonium ground-state spectrum compared to experimental
data.\label{fig:charmonium}}
\end{figure}
approach to hadron phenomenology via the corresponding inhomogeneous vertex BSEs, see e.g.,
\cite{Bhagwat:2007rj,Blank:2010sn,Blank:2011qk}.
Regarding the numerical solution of the quark DSE (\ref{eq:dse}) we note that in order to be able
to subsequently and consistently solve the BSE numerically, the propagator must be
known for quark four-momenta whose squares lie inside a parabola-shaped region of the
complex $p^2$ plane (for a more detailed discussion, see e.\,g., the appendix of Ref.~\cite{Krassnigg:2010mh}).
As a consequence, in particular for heavy quarks a reliable numerical approach to the quark DSE is
needed and we refer the reader to \cite{Krassnigg:2008gd} for the details of our particular solution method.
Furthermore, it is important to note here that the analytical structure of the quark propagator
can place restrictions in terms of an upper bound on the bound-state masses of mesons
accessible via standard numerical methods (see \cite{Bhagwat:2002tx} for a detailed discussion
and an initial step towards a full numerical treatment of such a situation). A route different
to the one described in \cite{Bhagwat:2002tx} is to extrapolate data obtained from the homogeneous BSE in the
accessible region to the inaccessible point of interest.
This is the approach we use here when necessary, and we give the details of our procedure in the appendix.

Once RL truncation has been chosen, the integral-equation kernels of (\ref{eq:bse}) and (\ref{eq:dse}) are
essentially characterized by an effective interaction $\mathcal{G}(s)$, $s:=(p-q)^2$.
The parameterization of Ref.~\cite{Maris:1999nt} reads
\begin{equation}\label{eq:interaction}
\frac{{\cal G}(s)}{s} = \frac{4\pi^2 D}{\omega^6} s\;\mathrm{e}^{-s/\omega^2}
+\frac{4\pi\;\gamma_m \pi\;\mathcal{F}(s) }{1/2 \ln [\tau\!+\!(1\!+\!s/\Lambda_\mathrm{QCD}^2)^2]}.
\end{equation}
This form produces the correct perturbative limit, i.\,e.~it preserves the one-loop renormalization
group behavior of QCD for solutions of the quark DSE. As given in \cite{Maris:1999nt},
${\cal F}(s)= [1 - \exp(-s/[4 m_t^2])]/s$, $m_t=0.5$~GeV,
$\tau={\rm e}^2-1$, $N_f=4$, $\Lambda_\mathrm{QCD}^{N_f=4}= 0.234\,{\rm GeV}$, and $\gamma_m=12/(33-2N_f)$.
The motivation for this function, which mimics
the behavior of the product of quark-gluon vertex and gluon propagator, is
mainly phenomenological. While currently debated on principle grounds (e.g.~\cite{Fischer:2008uz,Binosi:2009qm})
the impact of its particular form in the far IR on meson masses is expected to be small
(see also \cite{Blank:2010pa} for an exploratory study in this direction).

In \cite{Maris:1999nt}, together with the current-quark mass $m_q$, the parameters $\omega$ and $D$ were fitted to pion observables
and the chiral condensate. In this way, this effective coupling provided the correct amount of dynamical
chiral symmetry breaking as well as quark confinement via the absence of a Lehmann representation for
\begin{figure}
\includegraphics[width=\columnwidth,clip=true]{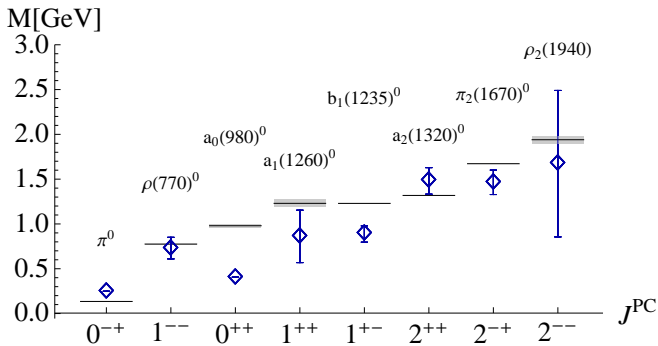}%
\caption{(Color online) The light-isovector ground-state spectrum compared to experimental
data.\label{fig:isovector}}
\end{figure}
the dressed quark propagator. Also, from the results of the fitting procedure in \cite{Maris:1999nt} it
was apparent that for a fixed current-quark mass one can obtain a good description of light
pseudoscalar and vector meson masses and decay constants by keeping the product
$D\times\omega=0.372$ GeV${}^3$ fixed and varying $\omega$ in the range $[0.3,0.5]$ GeV. In particular,
these observables were independent of $\omega$, which thus defines a one-parameter model. Later,
it was found that such a negligible dependence on $\omega$ is the characteristic of a ground state,
while exctiations---both radial and orbital---in general depend strongly on $\omega$ \cite{Holl:2004fr,Krassnigg:2009zh}.
As a result, it is possible to fix also $\omega$ to phenomenology, which we do in the present study,
albeit without fine-tuning the parameters any further, which is both besides the point
as well as beyond the scope of this work.

\section{Bottomonium\label{sec:bottomonium}}
With $D\times\omega$ fixed to the original value of $0.372$ GeV${}^3$ and taking into account the
trend visible in earlier computations of parts of the bottomonium spectrum e.g., in
\cite{Krassnigg:2010mh,Blank:2011qk}, we varied $\omega>0.5$ GeV and found good agreement
with all experimentally known bottomonium ground states at $\omega=0.61$ GeV via a least-squares fit of
the masses of the states with $J^{P}=0^-$, $0^+$, $1^-$, $1^+$, and $2^+$. The masses
for the states with $J^{P}=2^-$ are thus predictions of the model. The results
\begin{figure}
\includegraphics[width=\columnwidth,clip=true]{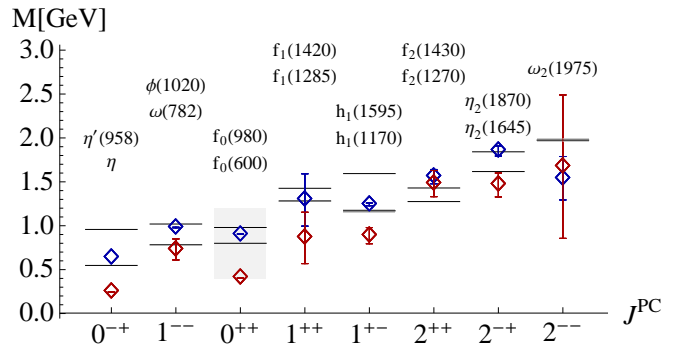}%
\caption{(Color online) The light-isoscalar ground-state spectrum compared to experimental
data. The red symbols correspond to the $n\bar{n}$-states of our ideally mixed
setup, the blue symbols denote pure $s\bar{s}$ states.\label{fig:isoscalar}}
\end{figure}
of our calculations are summarized in the first row of Tab.~\ref{tab:masses} and compared
to the available experimental data in Fig.~\ref{fig:bottomonium}. Our numerical uncertainties
are smaller than the sizes of the symbols in the figure for all cases.

These results for the masses show that already without fine-tuning we have achieved agreement on the level
of $3$ per-mill for all states considered, which is rather remarkable. In addition to
the excellent overall agreement we also achieve reasonable agreement with e.g., the experimental value of $69$ MeV for the
hyperfine splitting between the $0^{-+}$ and $1^{--}$ ground states, which is reproduced to 20\%.
\begin{table*}
\caption{Calculated meson masses in MeV (rounded) for the four quark-mass values
with extrapolation uncertainties given in brackets where applicable.
A comparison to experimental data is given in Figs.~\ref{fig:bottomonium} - \ref{fig:isoscalar}.
 Note that our results for $\bar{n}n$ correspond to both light isovector and isoscalar states.  \label{tab:masses}}
\begin{ruledtabular}
\begin{tabular}{rrrrrrrrr}
$J^{PC}$ & $0^{-+}$ & $1^{--}$ & $0^{++}$ & $1^{++}$ & $1^{+-}$ & $2^{++}$ & $2^{-+}$ & $2^{--}$ \\\hline
 $\bar{b}b$   & $9405$ & $9488$ & $9831$ & $9878$ & $9873$ & $9927$ & $10184(8)$ & $10188(8)$ \\
 $\bar{c}c$   & $2928$ & $3111$ & $3321$ & $3437$ & $3421$ & $3582$ & $3818(8)$ & $3818(9)$ \\
 $\bar{s}s$   &  $637$ & $980(3)$ & $904$ & $1293(299)$ & $1244(15)$ & $1560(81)$ & $1852(58)$ & $1541(247)$ \\
 $\bar{n}n$   &  $248$ & $730(120)$ & $405(1)$ & $861(292)$ & $886(92)$ & $1480(149)$ & $1465(136)$ & $1673(817)$ \\
\end{tabular}
\end{ruledtabular}
\end{table*}

Furthermore we computed the leptonic decay constants of the pseudoscalar and vector bottomonium states
and collect our results together with a comparison to experimental numbers (where available) in
Tab.~\ref{tab:dc}. While the experimental value for $f_{\eta_b}$ is yet unknown, we arrive
within 4\% of the experimental value for $f_\Upsilon$. This latter observation is remarkable
in particular, since the computation of the leptonic decay constants goes beyond spectroscopy
in that it involves also the BSAs of the states under investigation. This means
that the structure of the meson as described by the BSA is captured reasonably by
our setup as well.

\section{Charmonium and light mesons\label{sec:lower}}
While the main point of interest in the present work is the compatibility of our ansatz
with the ground state masses and decay constants of the bottomonium system, it is natural
to ask how the same parameter set performs for charmonium as well as strange and light
quark masses. We thus present the corresponding results in the remaining rows of
Tabs.~\ref{tab:masses} and \ref{tab:dc} and compare to available experimental data in
Fig.~\ref{fig:charmonium} for charmonium and Figs.~\ref{fig:isovector} and
\ref{fig:isoscalar} for states made out of  light and strange (anti)quarks.

\begin{table}
\caption{Calculated pseudoscalar and vector meson decay constants in MeV (rounded)
compared to experimental data \cite{Nakamura:2010zzi},
where available. \label{tab:dc}}
\begin{ruledtabular}
\begin{tabular}{rrl}
$J^{PC}$ & $0^{-+}$ & $1^{--}$  \\\hline
 $\bar{b}b$   & $708$ &  $687$  \\
 Experiment   & $-$   &  $715$  \\ \hline
 $\bar{c}c$   & $399$ &  $448$  \\
 Experiment   & $361$ &  $416$  \\ \hline
 $\bar{s}s$   & $173$ &  $237^{+4}_{-3}$  \\
 Experiment   & $-$   &  $227$   \\ \hline
 $\bar{n}n$ isovector & $108$ & $276^{+61}_{-113}$   \\
 Experiment   & $131$ &  $221$
\end{tabular}
\end{ruledtabular}
\end{table}
It is important to note here
that in our present truncation there is no flavor mixing, i.e., all states are
thus \emph{ideally mixed} and consequently \emph{a priori} can be expected to correspond
to experimental states only in the appropriate cases. Note also that we work in the
isospin-symmetric limit. In our tables we therefore list our results for pure $\bar{s}s$ and
$\bar{n}n$ states, where in the usual notation $n$ labels light quark flavors. While one
could apply simple flavor-mixing rules to our results \cite{Holl:2004un} we do not attempt this here
to maintain the clarity of our results as well as the simplicity of both the
model and the intention of our work. It is not aimed at a perfect description of light meson masses;
in fact, such an outcome cannot be expected of our present study, since RL truncation oversimplifies
the structure of the quark-gluon vertex for light and strange quarks (and apparently to some degree also
for charm quarks). However, it appears that a set of results such as the present one can in the future be
reconciled with experiment upon proper inclusion of corrections beyond RL truncation as they
have been explored in the recent literature, e.g., \cite{Fischer:2009jm,Chang:2009zb}.

Regarding our charmonium results we observe an overall pattern of agreement with experimental data
although some deficiencies are visible compared to the bottomonium case. Most notably the
scalar and axial-vector masses are underestimated while the hyperfine splitting is overestimated.
Our results for the pseudoscalar and vector leptonic decay constants are off $11$ and $8$\% of
the experimental numbers, which is about twice as much as for bottomonium.

For the strange and light quark cases we present our results compared to isoscalar and isovector
states as listed by the PDG \cite{Chang:2009zb}. Note that our results for $\bar{n}n$ as listed in Tab.~\ref{tab:masses}
correspond to both light isovector and isoscalar states. Due to the setup of the present study,
our observations here can reasonably only be of a general nature. For the isovector case one
observes a well-known pattern from the literature, namely that pseudoscalar and vector meson masses
compare well to experiment, while scalar and axial-vector states are substantially underestimated \cite{Krassnigg:2009zh}; for
tensor mesons the situation is better \cite{Krassnigg:2010mh}. Also for the decay constants, our model
provides a reasonable description of the data, even in its present form.

For the isoscalar case, the situation is naturally more complex. Still, the overall
impression is the same as described for the isovector case above except for the pseudoscalar
mesons, whose flavor composition cannot be described in RL truncation.

\section{Conclusions and outlook\label{sec:co}}
We have reported a study of the bottomonium ground-state meson masses
using and adjusting a well-established rainbow-ladder truncated
effective-interaction setup of the Bethe-Salpeter-equation. Our
goal was to provide---without fine-tuning---an initial test of
the potential such an approach holds to describe experimental data
in a reasonable fashion. Our results show that such a description
is indeed possible and beyond a surprisingly good match in bottomonium
also provides a reasonable description of charmonium and a
consistent picture for meson masses containing light and strange (anti)quarks.

This is an encouraging first step towards a comprehensive study
of mesons in this approach. Further steps will involve fine-tuning but
will also have to include appropriate corrections
beyond RL truncation to provide the mechanism for a reconciliation of
the deficiencies apparent in our present results for lower quark masses.
Together with an anchor of the model in the heavy-quark domain such as
the one exemplified here this will lead to a successful model description
of hadrons.

\begin{acknowledgments}
We acknowledge helpful conversations with D.~Horvati\'{c}, V.~Mader, C.~Popovici, and R.~Williams.
This work was supported by the Austrian Science Fund \emph{FWF} under project no.\ P20496-N16,  and
was performed in association with and supported in part by the \emph{FWF} doctoral program no.\ W1203-N08.
\end{acknowledgments}

\appendix*
\section{Technicalities}
As mentioned in Sec.~\ref{sec:equations} and described in detail in the
appendix of Ref.~\cite{Krassnigg:2010mh} the
Euclidean-space treatment of the meson BSE demands considerable conceptual and
numerical care. In particular, for the reasons given in Sec.~\ref{sec:equations}
one in some cases of higher-lying meson masses in our present approach has to
resort to extrapolation techniques. In \cite{Krassnigg:2010mh} a straight-forward
method was used, which we have since improved upon; our refined method is reported
in the following.

In the standard approach, the homogeneous BSE is solved as an eigenvalue
equation, where the on-shell point (and thus the mass of the meson in
question) is reached if the eigenvalue $\lambda(P^2=M^2)=1$ (for a detailed
discussion of BSE eigenvalues, see \cite{Blank:2010bp}). As
discussed in detail in \cite{Blank:2011qk}, this eigenvalue and its
dependence on the total-momentum squared is deeply connected to the
bound-state poles that appear in the corresponding four-point function,
\begin{equation}\label{eq:lambdapole}
\frac{\lambda(P^2)}{1 - \lambda(P^2)} = \frac{r}{P^2+M^2}+corrections \;,
\end{equation}
where $r$ denotes the residue at the pole corresponding to a particle of
mass $M$.

If the corrections in Eq.~(\ref{eq:lambdapole}) are neglected, the
combination $\frac{1 - \lambda(P^2)}{\lambda(P^2)}$ becomes linear in
$P^2$, which was used for the extrapolations done in
\cite{Blank:2011qk}. However, more reliable results can be obtained if
the corrections are taken into account. In this work, we assume
corrections of polynomial form,
\begin{equation}\label{eq:polcorr}
corrections = \sum_{i=1}^N \left( P^2\right)^i c_i\;,
\end{equation}
with constants $c_i$. These constants $c_i$, the residue $r$ as well as
the resulting masses $M$ are obtained in a straightforward fit to
calculated values of the function $\frac{\lambda(P^2)}{1 - \lambda(P^2)}$ in
the range of $P^2$ that can be accessed directly. In order to estimate the
uncertainty of the extrapolation, the polynomial order $N$ of the
corrections is varied from $N=1$ to $6$, and our final result is
computed from the arithmetic mean of this sample; the error bars are given
by the spread of the largest and smallest value.

A similar method is used fort the extrapolation of the decay constants $f$
of the light and strange vector states. In these cases, polynomials of degree
$3$, $4$, and $5$ have been fitted to the available values of $f(\sqrt{-P^2})\times\sqrt{-P^2}$
and extrapolated to $\sqrt{-P^2}=M$. Again, the average of the three resulting
values is quoted as final result, and the uncertainties are estimated from the
differences between the average and the largest and smallest value, respectively.

\end{document}